\documentclass[12pt,preprint]{aastex}

\newcommand{\vsini}{\mbox{$v\sin i$}}

\newcommand{\kms}{km s$^{-1}$}
\newcommand{\Teff}{$T_{\rm eff}$}
\newcommand{\logg}{$\log\thinspace g$}
\newcommand\vmac{\mbox{$v_{\rm mac}$}}
\newcommand\vmic{\mbox{$v_{\rm mic}$}}

\def\ltsima{$\; \buildrel < \over \sim \;$}
\def\simlt{\lower.5ex\hbox{\ltsima}}
\def\gtsima{$\; \buildrel > \over \sim \;$}
\def\simgt{\lower.5ex\hbox{\gtsima}}

\begin{document}

\title{First Magnetic Field Detection on a Class I Protostar\altaffilmark{1}}

\author{Christopher M. Johns--Krull}
\affil{Department of Physics \& Astronomy, Rice University, Houston, TX 77005}
\email{cmj@rice.edu}

\author{Thomas P. Greene}
\affil{NASA Ames Research Center, Moffett Field, CA 94035}
\email{thomas.p.greene@nasa.gov}

\author{Greg W. Doppmann}
\affil{NOAO, 950 North Cherry Ave., Tucson, AZ 85719}
\email{gdoppmann@noao.edu}

\author{Kevin R. Covey}
\affil{Harvard--Smithsonian Center for Astrophysics, Cambridge, MA 02138}
\email{kcovey@cfa.harvard.edu}

\altaffiltext{1}{Based on observations obtained at the Gemini Observatory, 
which is operated by the Association of Universities for Research in 
Astronomy, Inc., under a cooperative agreement with the NSF on behalf of the
Gemini partnership: the National Science Foundation (United States), the 
Science and Technology Facilities Council (United Kingdom), the National 
Research Council (Canada), CONICYT (Chile), the Australian Research Council
(Australia), Minist\'erio da Ci\^encia e Tecnologia (Brazil) and SECYT 
(Argentina).  The Phoenix data were obtained under the program: GS-2006A-C-12.}

\begin{abstract} 

Strong stellar magnetic fields are believed to truncate the inner accretion 
disks around young stars, redirecting the accreting material to the high 
latitude regions of the stellar surface.  In the past few years, observations 
of strong stellar fields on T Tauri stars with field strengths in general 
agreement with the predictions of magnetospheric accretion theory have 
bolstered this picture. Currently, nothing is known about the magnetic field 
properties of younger, more embedded Class I young stellar objects (YSOs).  
It is believed that protostars accrete much of their final mass during the 
Class I phase, but the physics governing 
this process remains poorly understood.  Here, we use high resolution near 
infrared spectra obtained with NIRSPEC on Keck and with Phoenix on Gemini 
South to measure the magnetic field properties of the Class I protostar 
WL 17.  We find clear signatures of a strong stellar magnetic field.  Analysis
of this data suggests a surface average field strength of $2.9 \pm 0.43$ kG on
WL 17.  We present our field measurements and discuss how they fit with the 
general model of magnetospheric accretion in young stars.

\end{abstract}


\section{Introduction} 

It is now generally accepted that accretion of circumstellar disk material
onto the surface of classical T Tauri stars (CTTSs) is controlled by
strong stellar magnetic fields (e.g. see review by Bouvier et al. 2007).
CTTSs represent Class II sources in the classification system defined by
Lada (1987).  The definition is based on a gradually falling spectral energy
distribution (SED) beyond $\sim 1 \mu$m.  This SED shape is believed to
arise from a geometrically thin, optically thick accretion disk containing
a high concentration of submicron sized dust grains (e.g. Bertout et al.
1988).  At some level, the final mass of these forming stars is determined by
how much of this disk material accretes onto the central star.  Additionally,
it is within the disks around these low mass pre-main sequence stars
that solar systems similar to our own form.   It is critical to
understand how the central young star interacts with and disperses its
disk in order to understand star, and particularly planet, formation.

The Class I sources defined by Lada (1987) represent one of the earliest
stages of star formation and are identified by a rising SED.  These sources
are deeply embedded within molecular clouds and are very faint or 
undetectable at optical wavelengths because of a thick envelope of 
circumstellar dust. It has been commonly thought that Class I objects 
represent an earlier evolutionary stage relative to Class II sources,
with the paradigm emerging that Class I sources are young protostars 
near the end of their bulk mass accretion phase.  This paradigm is bolstered
by the very weak photospheric absorption features in near infrared (IR)
spectra of these objects (e.g. Casali \& Matthews 1992; Greene \& Lada
1996, 2000).  The lack of absorption lines was interpreted by these
authors as the result of
strong veiling produced by emission originating in a vigorously accreting 
circumstellar disk which is being fed by an infalling envelope.  This 
emission is reprocessed by the dusty 
envelope which results in both the observed featureless continuum and the
rising SED.  As the accretion rate in the disk weakens and the thick 
circumstellar envelope either accretes onto the star plus disk
system or is disrupted
by strong outflows, it is generally thought that Class I objects evolve into
Class II sources.  

This general paradigm has been recently challenged by 
White \& Hillenbrand (2004) who find no strong differences in the properties
of the central stellar source between a sample of optically selected Class I
and Class II sources in Taurus.  On the other hand,
Doppmann et al. (2005) argue that the White \& Hillenbrand (2004) results are
biased by their optical selection of these Class I young stellar objects
(YSOs).  Doppmann et al. (2005)
perform an extensive IR study of Class I YSOs in several star forming 
regions and conclude that, while there is a fair amount of spread in the 
stellar and accretion properties of these objects, the general paradigm of
Class I sources representing an earlier, higher accretion rate phase of 
stellar evolution relative to Class II sources is borne out (see also
Prato et al. 2009).

Some of the confusion and disagreement over the true nature of the Class
I YSOs may be due to variability.  It has been suggested that the bulk
of a star's final mass is accreted through episodic events where the
accretion rate through the disk increases by a factor of $10 - 1000$ for
some period of time (e.g., Hartmann 1998).  These episodes of 
rapid disk accretion may be what we recognize as FU Orionis events
(Hartmann \& Kenyon 1996), with these events occuring more frequently
during the Class I stage.  As a result, Class I objects should display 
a large range of accretion behavior, with some objects accreting at close to
typical CTTS rates, while others are accreting much more rapidly than
this.  Qualitatively, such a picture matches the range of behavior found
in these sources in recent studies (Doppmann et al. 2005, White et al. 
2007, Prato et al. 2009).

Since Class I YSOs are often rapidly accreting material, the question
arises as to how this process occurs.  There is substantial evidence to show
that FU Ori outbursts are the result of very rapid disk accretion (for
a review see Hartmann
\& Kenyon 1996).  The evidence also suggests that when these objects are not
in outburst, accretion onto the central protostar occurs through a disk
with infalling material from the envelope piling up in the disk (e.g.
Bell 1993).  Such a
scenario can explain the apparent low luminosity of some Class I sources 
relative to what is expected if the infalling material from the envelope were
to land initially on the central object (Kenyon et al. 1993, 1994).  These
observations of FU Ori and lower luminosity Class I YSOs suggest that
accretion onto the central source occurs primarily through a 
disk whether a particular YSO is in a high or low accretion state.
For the Class II sources (CTTSs) this accretion process appears to be well 
described by the magnetospheric accretion paradigm (see Bouvier et al. 2007
for a review), but it is currently unclear to what extent this model
is appropriate for Class I protostars.

The magnetospheric accretion model is successful at explaining a number
of obsverations of CTTSs.  A key question in the study of these stars is 
to understand how they can accrete large amounts of disk material 
with high specific angular momentum, yet maintain rotation rates that are 
observed to be relatively slow (e.g. Hartmann \& Stauffer 1989, Edwards et 
al. 1994).  This problem is solved in current magnetospheric accretion models
by having the stellar magnetic field truncate the inner disk, typically near
the corotation radius, and channel the disk material onto the stellar 
surface, most often at high stellar latitude.  The angular momentum of the
accreting material is either transferred back to the disk (e.g., K\"onigl
1991; Cameron \& Campbell 1993; Shu et al. 1994) or is carried away by
some sort of accretion powered stellar wind (e.g., Matt \& Pudritz 2005).
Greene and Lada (2002) analyzed the stellar parameters and mass accretion
rate of the Class I source Oph IRS 43 and showed that these were consistent
with magnetospheric accretion models provided the magnetic field on this
source is on a order a kG in strength.  Covey et al. (2005) analyzed the 
rotational properties of Class I sources and found that while they are rotating
more rapidly than CTTSs on average, they are not rotating at breakup 
velocities.  These observations could be interpreted in the standard
magnetospheric accretion paradigm if the accretion rates of Class I sources
are larger on average than those of CTTSs.

Magnetospheric accretion naturally requires a strong stellar magnetic field.
Several TTSs have now been observed to have strong surface magnetic fields 
(Basri et al. 1992; Guenther et al. 1999; Johns--Krull 2007; Johns--Krull 
et al. 1999b, 2004; Yang et al. 2005, 2008), and strong magnetic fields have 
been observed in the formation region of the \ion{He}{1} emission line at 
5876 \AA\ (Johns--Krull et al.  1999a; Valenti \& Johns--Krull 2004; 
Symington et al.  2005; Donati et al. 2007, 2008), which is believed to be 
produced in a shock near the stellar surface as the disk material impacts 
the star (Beristain, Edwards, \& Kwan 2001).  While a considerable amount
is now known about the magnetic field properties of Class II YSOs, almost
nothing is known directly about the magnetic field properties of Class I
sources.  While not a Class I source, FU Ori has recently shown
evidence for a magnetic field in its disk, revealed through high resolution
spectropolarimetry (Donati et al. 2005); however, there are currently no
observations of magnetic fields on the surface of a Class I protostar.  This
is in part due to their faintness and the need for substantial observing
time on $8 - 10$ m class telescopes equipped with high resolution near
IR spectrometers in order to obtain the necessary data.

In order to begin to address the magnetic field properties of Class I
sources, we have begun an observational program to survey the magnetic
field properties of several Class I YSOs in the $\rho$-Ophiuchi star
forming region.  Here, we report on our first field detection on the
Class I source WL 17 (2MASS J16270677-2438149, ISO-Oph 103).  This
source has a rising IR SED (Wilking et al. 1989) with a spectral index of
$\alpha \equiv d{\rm log} \lambda F_\lambda / d{\rm log} \lambda = 0.61$
over the 2 -- 24 $\mu$m region (Evans et al. 2009).  WL 17 has been detected 
in X-rays (Imanishi et al. 2001; Ozawa et al. 2005) suggesting the 
star is magnetically active.  The temperature (\Teff$ = 3400$ K) and luminosity ($L_* = 1.8 L_\odot$) of WL 17 (Doppmann et al.
2005) give it a mass of $\sim 0.31 M_\odot$ and an age $\sim 10^5$ years
using the tracks of Siess et al. (2000).
In this paper, we look for Zeeman broadening of K-band \ion{Ti}{1} lines in high
resolution spectra of WL 17 to diagnose its magnetic field properties.  
Magnetic broadening is easiest to detect
when other sources of line broadening are minimized, and the small
rotation velocity of WL 17 ($v$sin$i = 12$ km s$^{-1}$; Doppmann et al.
2005) is a great advantage for this work.  Muzerolle et al. (1998) 
derive an accretion luminosity of $\sim 0.3 L_\odot$ based on the Br-$\gamma$
line luminosity measurements of Greene and Lada (1996).  
This accretion luminosity 
implies a mass accretion rate of $\dot M \sim 1.5 \times 10^{-7} 
M_\odot$~yr$^{-1}$ using equation (8) of Gullbring et al. (1998) with the 
disk truncation radius assumed to be at $5 R_*$.  While there are reasons
to be concerned about this mass accretion rate estimate (which are
discussed in \S 4), 
such a truncation radius
implies a stellar field of about a kilogauss.  Detecting and measuring that
field is the goal of the current work.  In 
\S 2 we describe our observations and data reduction.  The magnetic field 
analysis is described in \S3.  In \S 4 we give a discussion of
our results, and \S 5 summarizes our findings.

\section{Observations and Data Reduction}

Spectra analyed here for magnetic fields on WL 17 come from two sources.
Keck NIRSPEC data taken on UT 10 July 2001 is analyzed along with Gemini
Phoenix data.  The Keck data have already been published by Doppmann et al.
(2005) and the reader is referred to that paper for observing and data
reduction details.  For reference, the resolution of the NIRSPEC data is
$R \equiv \lambda / \delta \lambda \equiv 18,000$ (16.7~km~s$^{-1}$).
Spectra of WL~17 were acquired on UT 03 April 2006 with the Phoenix
near-IR spectrograph (Hinkle et al. 2002) on the 8-m Gemini South
telescope on Cerro Pachon, Chile.  Spectra were acquired with a
$0\farcs35$ (4-pixel) wide slit, providing spectroscopic resolution 
$R = 40,000$ (7.5~km~s$^{-1}$).  The
grating was oriented to observe the spectral range $\lambda$ =
2.2194--2.2300~$\mu$m in a single long-slit spectral order, and a slit
position angle of $90^{\circ}$ was used.  The seeing was approximately
$0\farcs50$ in $K$ band through light clouds, and WL~17 data were
acquired in two pairs of exposures of 1200 s duration each.  The
telescope was nodded $5\arcsec$ along the slit between integrations so
that object spectra were acquired in all exposures for a total of 80
minutes of integration time on WL~17.  The B1V star HR~5993 was
observed similarly but with shorter exposures for telluric correction
of the WL~17 spectra.  Both WL~17 and HR~5993 were observed at similar
airmasses, $X$ = 1.01 - 1.05.  Observations of a 
continuum lamp were acquired for flat fielding.

All data were reduced with IRAF. First, pairs of stellar spectra taken
at the two nod positions were differenced in order to remove bias, dark
current, and sky emission.  These differenced images were
then divided by a normalized flat field.  Next, bad
pixels were fixed via interpolation with the {\it cosmicrays} task,
and spectra were extracted with the {\it apall} task.  Spectra were
wavelength calibrated using low-order fits to 7 telluric absorption
lines observed in each spectrum, and spectra at each slit position
were co-added.  
Instrumental and atmospheric features were removed by dividing
wavelength-calibrated spectra of WL~17 by spectra of HR~5993 for each
of the two slit positions.  Final spectra were produced by combining
the corrected spectra of both slit positions and then normalizing the
resultant spectrum to have a mean continuum flux of 1.0.

\section{Analysis}

     The most successful approach for measuring magnetic fields on late--type 
stars in general has been to measure Zeeman broadening of spectral lines in
unpolarized light (Stokes $I$ e.g., Robinson 1980; Saar 1988; Valenti et al.
1995; Johns--Krull \& Valenti 1996; Johns--Krull 2007).  In the presence
of a magnetic field, a given spectral line will split up into a combination
of both $\pi$ and $\sigma$ components.  The $\pi$ components are linearly
polarized parallel to the magnetic field direction; and the $\sigma$ components
are circularly polarized when viewed along the magnetic field, and linearly
polarized perpendicular to the field when viewed from that direction.  The 
exact appearance of a line depends then on the details of the field strength
and direction, even when viewed in unpolarized light.  For any given Zeeman 
component ($\pi$ or $\sigma$), the splitting resulting from the magnetic field
is 
$$\Delta\lambda = {e \over 4\pi m_ec^2} \lambda^2 g B
                = 4.67 \times 10^{-7} \lambda^2 g B \,\,\,\,\,\, 
                {\rm m\AA},\eqno(1)$$
where $g$ is the Land\'e-$g$ factor of the transition, $B$ is the
strength of the magnetic field (given in kG), and $\lambda$ is the
wavelength of the transition (specified in \AA).

     Class I YSOs are relatively rapid rotators (e.g., Covey et al. 2005)
compared to most main sequence stars and most TTSs in which Zeeman broadening
has been detected, though WL 17 in particular has a relatively low $v$sin$i$. 
Equation (1) shows the broadening due to the Zeeman effect depends on the 
second power of the wavelength, whereas Doppler broadening due to rotation or
turbulent motions depends on wavelength to the first power.  Thus, observations
at longer wavelength are generally more sensitive to stellar magnetic fields. 
There are several \ion{Ti}{1} lines in the K band which are excellent probes
of magnetic fields in late-type stars (e.g. Saar \& Linsky 1985; Johns--Krull
2007), and here we observe 4 of them with NIRSPEC: (air wavelengths) 2.22112
$\mu$m with $g_{eff} = 2.08$, 2.22328 $\mu$m with $g_{eff} = 1.66$, 2.22740
$\mu$m with $g_{eff} = 1.58$, and 2.23106 $\mu$m with $g_{eff} = 2.50$.  We
observe the first 3 of these with Phoenix.  In addition to the
strongly Zeeman sensitive \ion{Ti}{1} lines, our wavelength settings also 
record a few additional atomic lines that are weaker and less Zeeman sensitive
(lower Land\'e-$g$ values) as well as the $v = 2-0$ CO bandhead at 
2.294 $\mu$m for the case of the NIRSPEC data.  
The CO lines are much less magnetically sensitive than the
\ion{Ti}{1} lines and provide a good check on other line broadening 
mechanisms.

In order to measure the magnetic field properties of WL 17, we directly 
model the profiles of several K band photospheric absorption lines.  
Our spectrum synthesis code and detailed analysis technique for measuring 
magnetic fields using these K band lines is described elsewhere (Johns--Krull
et al. 1999b; Johns--Krull et al. 2004, Yang et al. 2005).  Here, we simply 
review some of the specific details relevant to the analysis presented
here.  In order to synthesize the stellar spectrum, we must first specify
the atmospheric parameters: effective temperature (\Teff), gravity (\logg), 
metallicity ([M/H]), microturbulence (\vmic), macroturbulence (\vmac), and
rotational velocity (\vsini).  Rotational broadening in YSOs is large
compared to the effects of macroturbulence.  This makes it difficult to
solve for \vmac\ separately, so a fixed value of 2 \kms\ is adopted here as
it was in the above mentioned papers.  Valenti et al. (1998) found that 
microturbulence and macroturbulence were degenerate in M dwarfs, even with 
very high quality spectra.  Therefore, for the low turbulent velocities 
considered here, microturbulence is neglected, allowing \vmac\ to be a proxy
for all turbulent broadening in the photosphere.  While \vmic\ and \vmac\ can
in principle have different effects of the spectral lines (\vmic\ potentially
affecting the line equivalent width) at the relatively low resolution and
signal-to-noise used here, the effect of the two mechanisms on the {\it shape}
of the spectral lines is equivalent, and the corresponding broadening is
significantly less than that due to the resolution or magnetic fields.  Any
errors in the intrinsic line equivalent widths that result from an
inaccurate value of \vmic\ can in principle be compensated for by small
errors \Teff, \logg, [M/H], or the derived K band veiling.

With the turbulent broadening specified, estimates are still needed for \Teff,
\logg, \vsini, and [M/H] for WL 17.  Doppmann et al. (2005) find \Teff$ =
3400$ K and a gravity of \logg$ = 3.7$ for WL 17.  For the stellar atmosphere
then, we take the model from a grid of ``NextGen" atmospheres (Allard \& 
Hauschildt 1995) which is equal in effective temperature (3400 K) and closest
in gravity (\logg$ = 3.5$) to the values determined for WL 17.  Yang et al.
(2005), using the 4 \ion{Ti}{1} lines covered in the NIRSPEC data here,
performed several tests of the magnetic analysis methods used here
to see how sensitive the results are to small errors in the effective
temperature and gravity assumed in the magnetic analysis.  They find that
a 200 K error in \Teff\ or 0.5 dex error in \logg\ typically results in
less than a 10\% error in the derived magnetic field strength.  Therefore,
we are confident that our particular choice for the stellar atmosphere will
not lead to significant error in the magnetic field properties we estimate
for WL 17.  Finally, solar metallicity is often assumed for young stars, and 
this assumption is supported by the few detailed analyses that have been
performed (e.g. Padgett 1996).  We assume solar metallicity here for WL 17.
With the above quantities specified, we can then synthesize spectra for our
lines of interest using the polarized radiative code SYNTHMAG (Piskunov 1999).

The rotational broadening of WL 17 has been meaured by Doppmann et al. 
(2005), where they find \vsini$ = 12$ \kms.  The analysis of Doppmann et
al. (2005) used the CO bandhead data used here to measure \vsini.  Since
the CO lines are very insensitive to magnetic fields, we expect this to
be an accurate estimate of the rotational broadening of WL 17; however,
we let \vsini\ be a free parameter of our fits described below.  As 
mentioned above, Class I sources are often observed to have substantial K 
band veiling (e.g., Greene \& Lada 1996).  Veiling is an excess continuum
emission which, when added to the stellar spectrum, has the effect of 
weakening the spectral lines of the star in continuum normalized spectra.
Near infrared veiling is assumed to
arise from the disk around young stars.  Veiling is measured in units of the
local stellar continuum, and Doppmann et al. (2005) found
a K band veiling of $r_K = 3.9$ for WL 17 using the same NIRSPEC data we
use in part here.  Doppmann et al. (2003) showed that when magnetic fields
are unaccounted for in spectroscopic analysis, the results can be somewhat
biased.  Therefore, we choose to let the K band veiling be an additional
free parameter of the spectral fitting performed here.  In addition,
we attempt to simultaneously fit both Keck NIRSPEC and Gemini
Phoenix data which were observed at different times.  CTTSs regularly show
significant variations in their K band flux on timescales as short as a 
day (and occasionally shorter), likely as a result of accretion variability
(Skrutskie et al. 1996, Carpenter et al. 2001, Eiroa et al. 2002).  
Additionally, Barsony et al. 
(2005) have shown the the near-IR brightness of WL 17 is variable, suggesting
the K band veiling of WL 17 may be variable.  Therefore, we separately solve
for the K band veiling in the Keck and Gemini data.

In previous studies of the magnetic field properties of TTSs it was found
that the Zeeman sensitive \ion{Ti}{1} lines could not be well fit with a single 
value of the magnetic field strength.  Instead, a distribution of magnetic
field strengths provide a better fit (Johns--Krull et al. 1999b, 2004; 
Yang et al. 2005; Johns-Krull 2007).  
It was also found that fits to the spectra are 
degenerate in the derived field values unless the fit is limited to specific
values of the magnetic field strength, separated by $\sim 2$ kG, which is 
the approximate ``magnetic resolution" of the data.  While the NIRSPEC data
used here are slightly lower in resolution, the Phoenix data are a bit higher
in spectral resolution than that used in the studies cited above.  Therefore,
we use the same limitations when fitting the spectra of WL 17.  We assume the
star is composed of regions of 0, 2, 4, and 6 kG magentic field, and we solve
for the filling factor, $f_i$, of each of these components subject to the
constraint $\Sigma f_i = 1.0$.  The different 
regions are assumed to be well mixed over the surface of the star -- different
components are not divided up into well defined spots or other surface 
features.  The field geometry is assumed to be radial in all regions.
Another key assumption is
that the temperature structure in all the field regions is assumed to be 
identical for the purpose of spectrum synthesis:  the fields are not 
confined to cool spots or hot plage--like regions.  A final assumption we make
here is that the photospheric magnetic field properties of the star are the 
same between the
two observing epochs.  This may or may not be a good assumption.  Substantial
variability is seen in CTTSs, both photometrically and spectroscopically,
which has been interpreted as rotational modulation of a non axisymmetric
stellar magnetosphere (e.g., see Bouvier et al. 2007 for a review).  This 
certainly suggests variation of the field geometry above the star,
but not necessarily as much variation of the photospheric field itself.
Very little is known about variations over timescales of months to years
in the photospheric field 
properties of CTTSs (and nothing is known regarding Class I sources).  
Two field measurements exist for BP Tau (Johns--Krull et al. 1999b; 
Johns--Krull 2007) and for T Tau (Guenther et al. 1999; Johns--Krull 2007),
and for both stars, the mean field strengths recovered from the two epochs
agree to within the quoted uncertainties.  We therefore assume identical
field properties between the two epochs for WL 17 and show below that this
provides a good match to the data within the uncertainties. 

\begin{figure} 
\epsscale{.75}
\plotone{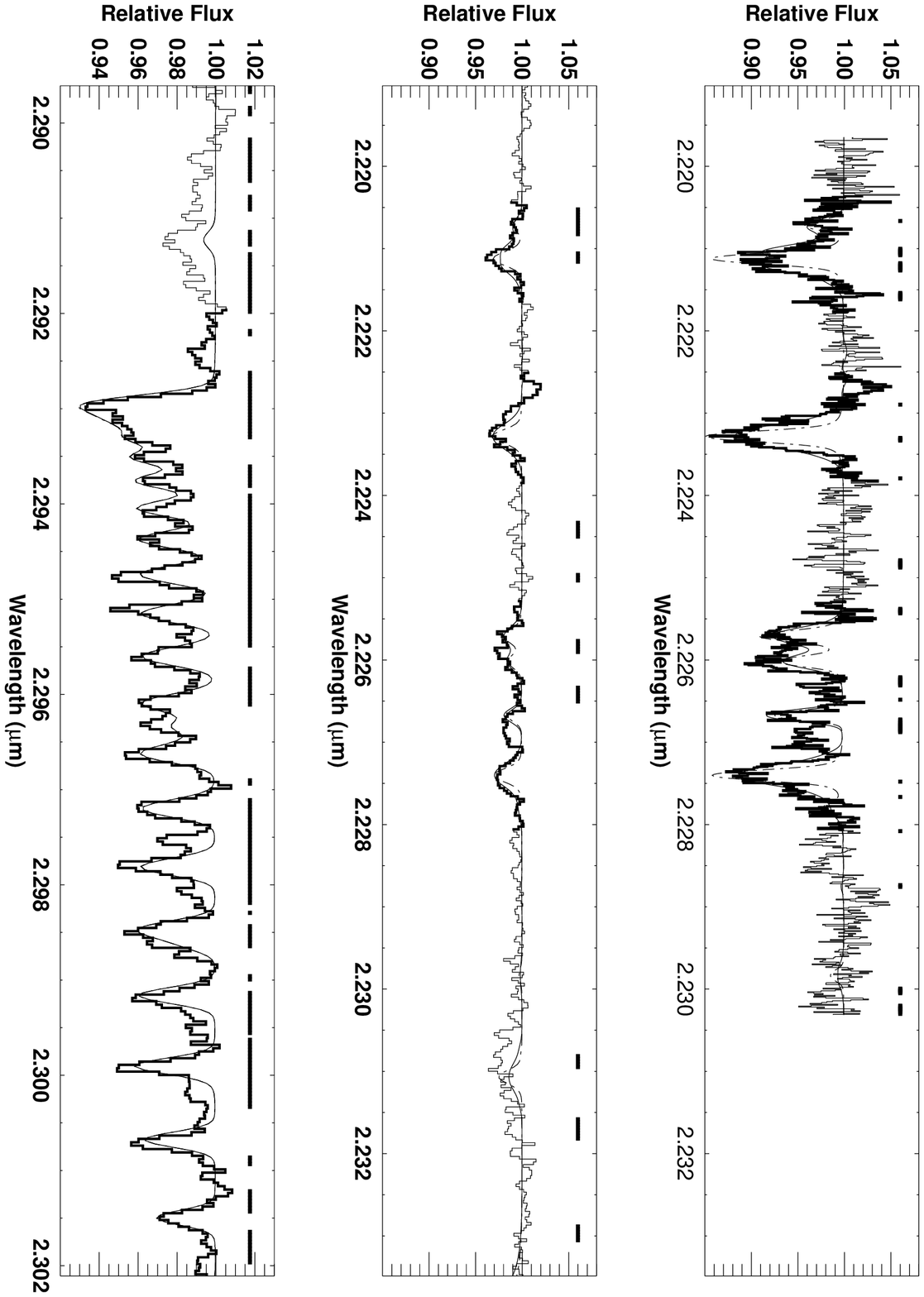}
\figcaption{Fits to the K band spectra of WL 17.  The observed spectra are
shown as the histogram in each panel.  The upper panel shows the Gemini
Phoenix data and the bottom two panels show the Keck NIRSPEC spectra.  Regions
where telluric absorption reached below 97\% of the continuum are
indicated by the thick lines above the spectra.  The bold histogram regions
show the parts of the spectra used in the third set of fits (F3) discussed
in the text.  In each panel, the fits including a magnetic field (from
fit F1) are shown in the smooth solid curve, and the best fit with no 
magnetic field is shown as the dash-dot curve.  The upper two panels are 
plotted on the same scale in
both axes to emphasize the change in K-band veiling between the two 
observing epochs.
\label{profiles}}
\end{figure}

There are then 6 free parameters in our model fits: the K band veiling, $r_K$,
for each observing epoch; the value of $f_i$ for the 2, 4 and 6 kG regions 
($\Sigma f_i = 1$, so the filling factor of the 0 kG field region is set once
$f_i$ is determined for the other 3 regions); and the value of \vsini.
Synthetic spectra are convolved with a Gaussian of the appropriate width
to represent the spectral resolution before comparison with the data.  We
have compared both calibration lamp lines and non-saturated telluric lines to
Gaussian fits and find that the line profiles are well matched by this assumed
line shape.  We are therefore confident that a Gaussian is a good approximation
for the instrumental profile.
We solve for our 6 free parameters using the nonlinear least squares 
technique of Marquardt (see Bevington \& Robinson 1992) to fit the model
spectra to the observed spectra shown in Figure \ref{profiles}.  In our
first attempt, labelled F1 in Table \ref{fitpars}, the entire
observed region shown in Figure \ref{profiles} is used in the fit.  
The parameters derived from all our fits are listed in Table \ref{fitpars}. 
In Figure \ref{profiles} we show the spectra of WL 17 in the regions of the
Zeeman sensitive \ion{Ti}{1} lines and the CO bandhead along with our best 
fitting model spectrum (F1).  Also included in the figure is a model with 
no magnetic field for comparison.  The Zeeman sensitive \ion{Ti}{1} lines 
at 2.2211, 2.2233, and 2.2274 $\mu$m (Land\'e-$g_{eff} = 2.08, 1.66, 1.58$, 
respectively) are significantly broader in the Phoenix data ($R = 40,000$) 
than is predicted by the model with no magnetic field.  On the other hand, 
the width of the Zeeman insensitive (Land\'e-$g_{eff} = 0.50$) \ion{Sc}{1} 
line at 2.2266 $\mu$m is accurately predicted by both models due to its much
weaker magnetic broadening.  This suggests that the excess broadening seen 
in the \ion{Ti}{1} lines is not due to an error in our assumed instrumental
profile.  In the lower resolution ($R = 18,000$) NIRSPEC data, the 
\ion{Ti}{1} lines again appear broader than predicted by the model with
no magnetic field, though the higher veiling associated with that data
makes the lines weaker which combined with the noise in the data makes
the reality of the excess broadening less certain than in the Gemini
Phoenix data.  However, the NIRSPEC data are fully consistent with the
magnetic broadening clearly seen in the Phoenix data.  The mean field, 
$\bar B = \Sigma B_i \times f_i$, that we find for WL 17 is 2.9 kG.

In the spectral regions used for this analysis there are some relatively
strong telluric absorption lines.  The spectra shown in Figure \ref{profiles}
have been corrected for telluric absorption; however, the regions affected
by this absorption are likely more uncertain than regions not affected by
such absorption.  In order to test the sensitivity of our results to 
possible errors in the telluric correction, we increased the uncertainty 
of spectral regions where the telluric absorption lines went below 97\% of 
the continuum during the fitting process.  These regions are
shown in the bold lines above the spectra in Figure \ref{profiles}.  In
these regions, we increased the uncertainty by a factor of 3 and 
reperformed the fit.  The fit parameters derived in this way are also
given in Table \ref{fitpars} as those derived from constraints F2.  The
values of the fitted parameters are almost identical to those derived
above (constraints F1).  A second concern is that the numerical fitting
method might be misled by small changes in $\chi^2$ due to the inclusion
of large amounts of continuum regions on which the model has no real
effect.  Therefore, we performed a third fit (labelled F3 in Table
\ref{fitpars}) in which we eliminated much of the continuum and focussed
down on the lines for fit constraints.  The regions of the spectra
used for this fit are shown in bold in Figure \ref{profiles}, and the
fit parameters are again reported in Table \ref{fitpars}.  For this
fit, we maintained the uncertainty of the telluric affected regions at
3 times their nominal values.  Again, the fitted parameters are nearly 
identical to those determined in fits F1 and F2.

\section{Discussion}

Our detection of a mean field of $\bar B = 2.9$ kG on WL 17 is the first
magnetic field measurement on a Class I protostar.  Previous studies using K
band data of comparable resolution and signal-to-noise level have shown 
that the field uncertainties are dominated by systematic effects associated
with the choice of magnetic model (Johns--Krull et al. 1999b, 2004; Yang et
al. 2008) and possible errors in the stellar parameters used to model the star
(Yang et al.  2005).  Based on these studies, while the formal uncertainty in
the mean field determination is quite low, we estimate the true uncertainty
in our mean field measurement of be
$\sim 10-15$ \%.  Johns--Krull (2007) measured the
mean magnetic field on a sample of 14 CTTSs in the Taurus star forming 
region, finding field strengths which ranged from 1.1 to 2.9 kG, with a
mean of 2.2 kG.  Yang et al. (2005, 2008) measured the mean field on a
total of 5 stars in the TW Hydrae association (TWA) finding values that range
from 2.3 to 3.5 kG with a mean field of 3.0 kG.  Yang et al. (2008) find
that this difference in mean field strength between the two samples is
marginally significant, with the older stars (TWA) having a larger field
strength on average.  However, Yang et al. (2008) point out that the TWA
stars have smaller radii on average due to their older age ($\sim 10$ Myr 
compared to $\sim 2$ Myr for Taurus), and that the mean magnetic flux in
the TWA stars is actually smaller than that in the Taurus stars.
WL 17 has a field
strength that is large relative to the Taurus stars studied by Johns--Krull
(2007) and it also has a relatively large radius and corresponding high
magnetic flux.  On the other hand, WL 17 is in many ways similar to
DF Tau, which has both a large radius (3.4 $R_\odot$) and a mean field of
2.9 kG, equal to that of WL 17.  Observations of a statistically significant
sample of Class I sources will be required to see how their magnetic field
properties compare as a group to older populations of Class II and III 
(diskless T Tauri stars) objects.

Our derived \vsini$ = 11.7$ km s$^{-1}$, 
with a formal uncertainty of $0.4$ km s$^{-1}$, is consistent with the value of
12 km s$^{-1}$ reported by Doppmann et al. (2005).  The veiling we derive for
WL 17 is quite different in the two epochs.  For the NIRSPEC data, we find
$r_K = 6.4 \pm 0.1$.  Using the same NIRSPEC data set, Doppmann et al. (2005)
quote a veiling value of $r_K=3.9$, which includes a correction for a 
systematic effect seen in the best fit synthesis models to observations of 
MK standards.  The measured veiling from Doppmann et al. (2005)
without the correction for the systematic
effect was $r_K=4.9$, based on fits to two wavelength regions containing
strong lines of Al, Mg, and Na.  The CO bandhead was the third wavelength 
region used in  the Doppmann et al. (2005) study, but was only used to derive
the \vsini\ rotation value.  As a result, there are actually no wavelength
regions in common between this study and that of Doppmann et al. (2005) for
the purpose of determining $r_K$.  We do note that when using only the CO
bandhead region of WL 17, Doppmann et al. find a value of $r_K = 7.1$ 
(Doppmann, private communication), though this region was not actually used
in their final veiling determination.  

Another difference between this study and that of Doppmann et al. (2005) for
the determination of the veiling is the inclusion of magnetic fields.  All the
atomic lines used in this analysis and that of Doppmann et al. have some
Zeeman sensitivity.  At the resolution of the NIRSPEC data ($R = 18,000$)
and for the strong, broad (e.g. \ion{Na}{1}) lines used by Doppmann et al. 
(2005), magnetic fields primarily increase the equivalent width of the lines
compared to models which do not include fields (Doppmann et al. 2003).  As a
result, the somewhat stronger lines produced by magnetic models must be diluted
by more veiling flux than that required for these same (weaker) lines 
produced by non-magnetic
models in order to match a given set of observed line strengths.  The
veiling, $r_K$, inferred from an observed spectrum will thus be larger when
derived by comparison to magnetic models and smaller when derived by 
comparison to non-magnetic models.  This is the effect we see when 
comparing the results here with those of Doppmann et al. (2005).
Our veiling estimate of $r_K = 6.4$ from NIRSPEC is a little less than
Doppmann et al. find from the CO bandhead alone, but a little larger than
the $r_K = 4.9$ they find from their atomic lines, though accounting for
magnetic fields would likely bring their $r_K = 4.9$ value up by some amount.
We note that we use a single value of $r_K$ to fit both the CO bandhead and 
the \ion{Ti}{1} line region in the NIRSPEC data (lower two panels of Figure
\ref{profiles}).  Some of the difference in $r_K$ found by Doppmann et al.
(2005) in different wavelength regions is likely due to model atmosphere
and line data uncertainties (see also a discussion of
this in Doppmann et al. 2005).  As a result, it is likely that our formal
uncertainty of 0.1 on the veiling is too low, so we arbitrarily increase it
by a factor of 3 and estimate a veiling for our NIRSPEC data of
$r_K = 6.4 \pm 0.3$.  We adopt the same uncertainty for our Phoenix
data, which is at a comparable signal-to-noise, giving $r_K = 1.1 \pm 0.3$ 
for this observation time.  

Assuming the veiling differences quoted above result in a
corresponding change to the K band
source brightness, and that only the strength of the veiling continuum
changed between these observations (i.e. that the underlying star remained
constant), WL 17 should have been a factor of 
$(1.0 + 6.4 \pm 0.3)/(1.0 + 1.1 \pm 0.3) = 3.5 \pm 0.5$
brighter in the K band at the time of the Keck NIRSPEC observation relative 
to the Gemini Phoenix observation.  Interestingly, if we adopt the K band
veiling correction of Doppmann et al. (2005), the individual veilings for
each epoch are lowered, but the predicted flux ratio between the two
observing epochs remains almost the same (3.4 instead of 3.5).  

     The K band flux factor variation of 3.5 calculated above corresponds 
to a variation of 1.36 magnitudes.  There are relatively few studies of the
near-IR variability of Class I sources; however, the few that exist suggest
that while the implied K band variability of WL 17 is large, it may not be
too extreme for such a source.  Kenyon and Hartmann (1995) plot a histogram
of the standard deviation for protostars in their sample that have two
or more K band photometric measurements, finding values that reach as high
as $\sigma_K \sim 0.8$.  In their study, Park and Kenyon (2002) find values
of $\sigma_K$ up to 0.52.  Our two veiling meaurements give $\sigma_K = 0.96$,
while Barsony et al. (1997) report $\sigma_K = 0.57$ for WL 17 on the basis
of 6 different K band photometric measurements.
Barsony et al. (2005) tabulate $r_K$ variations for several sources in 
$\rho$ Oph, and while none show quite as large a variation as we find for
WL 17, a few other show large $r_K$ variations with some ranging in value from
1 -- 4.  Barsony et al. (2005) also study the mid-IR variability of their
sources, finding that WL 17 varies in 12.5 $\mu$m flux by a factor of 2.4
which is similar in magnitude to the factor of 3.5 change in the K band 
brightness found here.  Values of $r_K \geq 0.5$ are usually taken to
indicate active accretion from a circumstellar disk, and large variable
K band veiling such as that shown by WL 17 and several other sources in
$\rho$ Oph is usually taken as evidence that this accretion can be highly
time variable (e.g. Barsony et al. 2005).  The exact cause of this high
degree of variability is not yet clear however.

The veiling analysis and implied K band photometric variations described
above assumes the underlying star remains constant; however, it is likely
that the star also possesses cool starspots which could contribute some K
band variation due to rotational modulation.  Veiling is usually attributed
to a source producing a featureless continuum.  Starspots themselves contribute
very weakly to veiling variations since the spectrum of the spots contains
many of the lines in present in the non-spotted photosphere.  The potential
effect of this can be estimated by measuring the veiling of a non-accreting
T Tauri star using another non-accreting star as a template.  In the
optical, this level of veiling is $\sim 0.1$ (Hartigan et al. 1991) and
should be smaller in the K band since the spot quiet atmosphere contrast 
is much weaker.  Weakly or non-accreting T Tauri stars do
show K band brightness variations with a peak-to-peak amplitude $\leq 0.2$ 
magnitudes as a result of spots (e.g. Skrutskie et al. 1996, Carpenter et al.
2001) which is substantially less than the K band variations implied by
our veliling measurements or observed by Barsony et al. (2005).  Thus, 
the majority of the K band veiling variation here must be due to changes in
the accretion properties of WL 17.

One of the motivations for this study is to see to what extent the 
magnetospheric accretion paradigm in place for Class II YSOs may be applicable
to Class I sources.  Johns--Krull et al. (1999b) give equations for predicting
the stellar magnetic field strength required for 3 different prescriptions
(K\"onigl 1991, Cameron \& Campbell 1993, Shu et al. 1994) of magnetospheric
accretion theory.  The work of K\"onigl (1991) and Shu et al. (1994, see also
Camenzind et al. 1990) give the same scaling with stellar and accretion
parameters: $B_* \propto M_*^{5/6} R_*^{-3} P_{rot}^{7/6} \dot M^{1/2}$.
The scaling from Cameron and Campbell (1993) is very similar:
$B_* \propto M_*^{2/3} R_*^{-3} P_{rot}^{29/24} \dot M^{23/40}$.
Using these equations to predict the field on WL 17 is 
uncertain due to difficulties with estimating the luminosity of WL 17 (which
affects the derived mass, radius, and accretion rate).  Luminosity estimates
for WL 17 range from 1.8 $L_\odot$ (Doppmann et al. 2005) to 0.12 $L_\odot$
(Bontemps et al. 2001).  Additional uncertainties also affect the mass 
accretion rate, and the rotation period of WL 17 is unknown.   Assuming 
\Teff$=3400$ K from Doppmann et al. (2005) is
a fairly robust estimate, we use this in combination with the two
quoted stellar luminosities to derive the quantities needed to predict the
stellar field strength from magnetospheric accretion theory.  These stellar
properties are reported in Table \ref{magaccretion} along with the field 
predictions for the studies mentioned above.  We note that the stellar
luminosity from Bontemps et al. (2001) in combination with the 3400 K 
effective temperature would give WL 17 an age of $\sim 5$ Myr using the 
pre-main sequence tracks of Siess et al. (2000).  Such an age would
be unusual for a Class I source.  
The Doppmann et al. (2005) stellar luminosity may be more accurate than
the Bontemps et al. (2001) value. This is because the former was derived
by using photometric measurements to estimate and correct for the
extinction and veiling seen toward the photosphere of the
spectroscopically determined effective temperature, while the latter was
determined by de-reddening near-IR photometry of WL 17 to intrinsic CTTS
colors and assuming an intrinsic J band flux to stellar luminosity
relationship for a typical CTTS. The Doppmann et al. (2005) approach
corrects for the veiling and effective temperature measured specifically for
WL 17, while the Bontemps et al. (2001) technique does not.  Nevertheless,
we report magnetospheric accretion estimates based on both values of the
stellar luminosity in order to illustrate the sensitivity of the expected
fields to various stellar parameters, notably the luminosity.

In addition to issues related to the correct luminosity for
WL 17, there is additional uncertainty regarding the accretion rate estimate.
The value of $\dot M \sim 1.5 \times 10^{-7} M_\odot$~yr$^{-1}$ quoted in
\S 1 is based on the accretion luminosity estimate of Muzerolle et al.
(1998) which is in turn based on the Br-$\gamma$ line luminosity estimate
from Greene and Lada (1996).  A major concern in this process is the
extinction correction used to recover the Br-$\gamma$ line luminosity.  
For example, Greene and Lada (1996) corrected their data for WL 17
back to the CTTS locus in the JHK color--color diagram, 
not necessarily back to the stellar photosphere.
As an example of the sensitivity to the details of extinction corrections
and the photometric data used, we note that Doppmann et al. (2005) also 
compute Br-$\gamma$ line luminosities (their Figure 11).  These authors
de-redden the H-K color to an intrinsic value of 0.6 and then add in a 
correction for scattered light based on the models of Whitney et al. (1997).  
This results in a Br-$\gamma$ line luminosity of $6.8 \times 10^{-4} L_\odot$
(Doppmann, private communication),
which in turn gives an accretion luminosity of 2.7 $L_\odot$ using the
Muzerolle et al. (1998) relationship.  Calculating the mass accretion rate
as given in the introduction results is a value of $\dot M \sim 1.4 \times
10^{-6} M_\odot$~yr$^{-1}$.  We include the field estimates resulting from
this accretion rate in Table \ref{magaccretion}, and we note that such an
accretion rate  is about the level needed to accrete a 0.5 $M_\odot$ star in 
a $\sim 3.5 \times 10^5$ yr.  This large accretion rate estimate for a Class I
source is also supported by the estimate of 
$\dot M = 1 \times 10^{-6} M_\odot$~yr$^{-1}$ found by Greene and Lada (2002) 
for Oph IRS 43.

Obviously, the accretion rate and implied magnetic field are fairly
sensitive to the details of the Br-$\gamma$ line luminosity calculation.
We therefore recomputed the Br-$\gamma$ line luminosity from the measured
equivalent width value of 4.3 \AA\ (Doppmann et al. 2005), the 2MASS
photometry (Skrutskie et al. 2006), and correcting for extinction by 
de-reddening the JHK colors to the CTTS locus and correcting for an extra 
$A_K = 0.88$ mag (see Doppmann et al. \S 3.8). We also used a distance of 
135 pc to the $\rho$ Oph cloud (Mamajek 2008). This produced a Br-$\gamma$
line luminosity of $2.9 \times 10^{-4} L_\odot$, about 2.5 times higher than
that given by Greene \& Lada due mostly to the more recent photometry and 
extinction correction technique. This new line luminosity indicates an 
accretion luminosity of 0.9 $L_\odot$ from the Muzerolle et al. (1998) 
relationship, which implies a mass accretion rate of $4.5 \times 10^{-7}
M_\odot$~yr$^{-1}$, which is well bounded by the accretion rates given in
Table \ref{magaccretion}.

All of the above discussion implicitly assumes the Muzerolle et al. (1998) 
Br-$\gamma$ accretion luminosity relationship holds for Class I sources;
however, this relationship was derived based on a sample of Class II objects.
If there are any systematic differences between these and Class I 
sources, the resulting accretion rate will be in error.  For example, if
the Br-$\gamma$ line is more optically thick in Class I sources due to
systematically higher accretion rates, the accretion rate we derive above
for WL 17 will be lower than the true value.  
This suggests that the derived accretion rates for WL 17 may be too low.

  The rotation period reported 
in Table \ref{magaccretion} is an upper limit based on the stellar radius 
and measured \vsini.  If the rotation period is actually shorter, the derived
magnetic field values will be smaller.  In this sense,
the values reported in the Table
are an upper limit depending on the inclination of the source.  The field
strengths reported in Table \ref{magaccretion} are those corresponding to
the equatorial field strength for an assumed dipolar field geometry.  The
polar field strength in such a geometry is twice this value.  The mean field
of 2.9 kG found for WL 17 is well above the predicted values for the larger
luminosity found by Doppmann et al. (2005); while for the lower luminosity of
Bontemps et al. (2001), the measured field value may not be strong enough,
particularly if the field is not dominated by the dipole component.  In
most TTSs, the dipole component is found to be weak
(Johns--Krull et al. 1999a; Daou et al. 2006;
Yang et al. 2007; Donati et al. 2007).  It is important to note that the
data presented here only probe the photospheric field strength, while 
providing few constraints on the field geometry.  High spectral resolution
near-IR circular spectropolarimetry will likely be required to explore the 
field geometry on Class I YSOs such as WL 17.

In summary, the magnetic field we measure on WL 17 is roughly consistent 
with predicted magnetic field strength required for magnetospheric 
accretion.  However, detailed quantitative comparisons are greatly
hampered due to a number of uncertainties related to other relevant
stellar parameters that are currently difficult to estimate for Class
I sources (see also the discussion in Prato et al. 2009).  Class I
sources by their nature are deeply embedded and therefore suffer 
substantial extinction.  Uncertainties involved with the proper way
to de-redden observations of these sources, combined with variable
accretion luminosity (continuum and line), a general lack of measured
rotation periods for Class I objects, and uncertainties in the methods
used to measure accretion rates suggest that much work is still left
to do before the magnetospheric accretion paradigm can be firmly
established or refuted for Class I sources.

On the Sun and active stars, it is expected that magnetic flux tubes are
confined at photospheric levels by the gas pressure in the external 
non-magnetic atmosphere.  For example, Spruit \& Zweibel (1979) computed flux
tube equilibrium models, showing that the
magnetic field strength is expected to 
scale with gas pressure in the surrounding non-magnetic photosphere.  Similar
results were found by Rajaguru et al. (2002).  Field strengths set by such 
pressure equipartition considerations appear to be observed in active G and K
dwarfs (e.g. Saar 1990, 1994, 1996) and possibly in M dwarfs (e.g. 
Johns--Krull \& Valenti 1996).  Class I YSOs have relatively low surface 
gravities and hence low photospheric gas pressures, so that equipartition 
flux tubes would have relatively low magnetic field strengths compared to 
cool dwarfs.  The maximum field strength allowed for a confined magnetic 
flux tube is $B_{eq} = (8 \pi P_g)^{1/2}$ where $P_g$ is the gas pressure at
the observed level in the stellar atmosphere.  Here, we take as a lower limit
in the atmosphere (upper limit in pressure) the level where the local 
temperature is equal to the effective temperature (3400 K) in the NextGen
models of the appropriate gravity.  This is the 
approximate level at which the continuum forms, with the \ion{Ti}{1} lines 
forming over a range of atmospheric layers above this level at lower
pressure.  The values of $B_{eq}$ are given in Table \ref{magaccretion}.
The mean field we measure for WL 17 is well above the value of $B_{eq}$ for
either assumed luminosity and resulting gravity.  This suggests that pressure
equipartition does not hold in the case of Class I YSOs, and it also suggests
the gas pressure in the atmospheres of these young stars is dominated by 
their magnetic fields.  Indeed, our fit for the magnetic field of WL 17
has only 3\% of the surface as field free; however, the uncertainty on the
filling factor of this component
is such that this is not a significant measurement.  It may
well be that the entire surface of WL 17 is covered with strong magnetic 
fields.  What is certain is that the field we measure is too strong to be 
accounted for by current models of dynamo action on fully convective stars, 
where the field strength is equal to the equipartition value (Bercik et al. 
2005, Chabrier \& K\"uker 2006, Dobler et al.  2006) and the filling factor of
this field is typically very low (Cattaneo 1999, Bercik et al. 2005).
Rapid rotation can enhance field production and may help organize magnetic
fields into large scale dipolar or quadrapolar geometries (e.g. Dobler et al.
2006, Brown et al. 2007); however, it is not yet clear that such models 
can produce mean fields on young stars in the range of $2-3$ kG.

Stellar magnetic fields give rise to ``activity" in late-type stars.  
Activity is typically traced by line emission or broad band emission at 
high energy wavelengths such as X-rays.  Pevtsov et al. (2003) 
find an excellent correlation between the X-ray luminosity, $L_X$, and 
magnetic flux in solar active regions and dwarf type stars with each of
these two quantities ranging over almost 11 orders of magnitude.
Using the X-ray luminosity--magnetic flux relationship of Pevtsov et al. 
(2003) with our mean magnetic field measurement of 2.9 kG for WL 17,
we can predict $L_X$ for this YSO if the radius is known.  In Table
\ref{magaccretion} we report the predicted X-ray luminosity values for
the two stellar radii implied by the different stellar luminosity 
estimates of Bontemps et al. (2001) and Doppmann et al. (2005).  WL 17
has been detected in X-rays by Chandra (Imanishi et al. 2001) and by
XMM (Ozawa et al. 2005), and possibly by ASCA, though crowding was an
issue for this latter observation (Kamata et al. 1997).  The Chandra
observation gives an X-ray luminosity of $L_X = 3.1 \times 10^{29}$
erg s$^{-1}$ while the XMM observation found $L_X = 9.6 \times 10^{29}$
erg s$^{-1}$.  Both values are lower than either of the predicted values
in Table \ref{magaccretion}, though the XMM value of Ozawa et al. (2005)
is not much lower than the lower estimate which is based on the Bontemps
et al. (2001) luminosity for WL 17.  We do note though that the Pevtsov
et al. (2003) X-ray luminosity--magnetic flux relationship is based on
observed X-ray emission outside obvious flares: the so-called quiescent 
emission level.  The XMM X-ray lightcurve of WL 17 presented by Ozawa
et al. (2005) is dominated by a strong, long lasting flare.  Thus, the
quiescent X-ray emission of WL 17 is likely significantly weaker than
the predicted X-ray emission, independent of the exact value of the
the stellar radius for WL 17.  Pevtsov et al. (2003) found evidence that
TTSs were underluminous in X-rays relative to the relationship defined by
dwarf stars, and this result appears to hold for a majority of these stars
(Johns--Krull 2007; Yang et al. 2008).  In that sense, WL 17 follows the
same pattern seen in most Class II sources.  Johns--Krull (2007) suggested
that this might be due to the fact that the magnetic pressure at the surface
of these stars is so strong.  Perhaps the ubiquitous, strong fields of
these young stars inhibit (relative to dwarf stars) photospheric gas motions
from moving flux tubes around and building up magnetic stresses which 
eventually results in coronal heating.  Alternatively, WL 17 is deeply
embedded and may suffer substantial X-ray attenuation, though this in
nominally accounted for in the X-ray measurements.

\section{Summary}

We have measured the magnetic field on the Class I source WL 17, and we
find the star is largely covered by strong magnetic fields.  The surface
averaged mean field on the star is $2.9 \pm 0.43$ kG.  Comparing this
field with predictions from magnetospheric accretion theory or with
X-ray measurements depends fairly sensitively on the value of the stellar
radius appropriate for WL 17.  For the relatively large luminosity and
associated stellar radius from Doppmann et al. (2005), the measured field
values are more than strong enough to be consistent with magnetospheric
accretion; however, for the lower radius implied by the Bontemps et al.
(2001) luminosity, the field may in fact be too weak.  For either radius,
the measured fields are stronger than expected by pressure balance arguments
for magnetic flux tubes confined by a non-magnetic atmosphere, suggesting
that the star is likely fully covered by fields.  In addition, independent
of the radius assumed, the measured X-ray emission is lower than is to 
be expected based on correlations established between these two quantities
on the Sun and late-type dwarf stars.

\acknowledgements
We are pleased to acknowledge numerous, stimulating discussions with J. Valenti
on all aspects of the work reported here.  We also acknowledge many useful
comments and suggestions from an anonymous referee. CMJ-K wishes to acknowledge 
partial support for this research from the NASA Origins of Solar Systems 
program through grant numbers NAG5-13103 and NNG06GD85G made to Rice 
University.  TPG acknowledges support from NASA's Origins of Solar Systems 
program via WBS 811073.02.07.01.89.  KRC acknowledges support for this work 
by NASA through the Spitzer Space Telescope Fellowship Program, through a 
contract issued by the Jet Propulsion Laboratory, California Institute of 
Technology under a contract with NASA.
This work made use of the SIMBAD reference database, the NASA
Astrophysics Data System, and VALD line database.

\clearpage

%
\clearpage
 
\begin{deluxetable}{lcccccccc}
\tablewidth{14.5truecm}   
\tablecaption{Fit Parameters\label{fitpars}}
\tablehead{
   \colhead{}&
   \colhead{$v$sin$i$}&
   \colhead{$r_K$}&
   \colhead{$r_K$}&
   \colhead{}&
   \colhead{}&
   \colhead{}&
   \colhead{}&
   \colhead{$\Sigma B_i \times f_i$}\\[0.2ex]
   \colhead{Fit}&
   \colhead{(km s$^{-1}$)}&
   \colhead{(NIRSPEC)}&
   \colhead{(Phoenix)}&
   \colhead{$f_{\rm 0 kG}$}&
   \colhead{$f_{\rm 2 kG}$}&
   \colhead{$f_{\rm 4 kG}$}&
   \colhead{$f_{\rm 6 kG}$}&
   \colhead{(kG)}
}
\startdata
F1 & 11.7 & 6.4 & 1.1 & 0.03 & 0.60 & 0.23 & 0.14 & 2.9 \\
F2 & 12.0 & 6.4 & 1.0 & 0.08 & 0.55 & 0.21 & 0.16 & 2.9 \\
F3 & 11.7 & 6.5 & 1.0 & 0.03 & 0.64 & 0.17 & 0.16 & 2.9 \\
\enddata
\end{deluxetable}

\clearpage

\begin{deluxetable}{lccccccccccc}
\tablewidth{17.5truecm}   
\tablecaption{Magnetospheric Accretion Predictions\label{magaccretion}}
\tablehead{
   \colhead{\Teff}&
   \colhead{$L_*$}&
   \colhead{$R_*$\tablenotemark{a}}&
   \colhead{$M_*$\tablenotemark{b}}&
   \colhead{$P_{rot}$\tablenotemark{c}}&
   \colhead{$\dot M$\tablenotemark{d}}&
   \colhead{$B$\tablenotemark{e}}&
   \colhead{$B$\tablenotemark{f}}&
   \colhead{$B$\tablenotemark{g}}&
   \colhead{\logg}&
   \colhead{$B_{eq}$}&
   \colhead{$L_X$\tablenotemark{h}}\\[0.2ex]
   \colhead{(K)}&
   \colhead{($L_\odot$)}&
   \colhead{($R_\odot$)}&
   \colhead{($M_\odot$)}&
   \colhead{(days)}&
   \colhead{($M_\odot yr^{-1}$)}&
   \colhead{(kG)}&
   \colhead{(kG)}&
   \colhead{(kG)}&
   \colhead{(cgs)}&
   \colhead{(kG)}&
   \colhead{($10^{30}$ erg s$^{-1}$)}
}
\startdata
3400&1.8&3.9&0.31&16.9&$1.5 \times 10^{-7}$&0.72&0.33&0.85&2.7&1.06\tablenotemark{i} & 29.0 \\
3400&0.12&1.0&0.29&4.33&$4.3 \times 10^{-8}$&4.43&1.74&5.22&3.9&1.44\tablenotemark{j} & 1.3 \\
3400&1.8&3.9&0.31&16.9&$1.4 \times 10^{-6}$&2.20&1.18&2.60&2.7&1.06\tablenotemark{k} & 29.0 \\
\enddata
\tablenotetext{a}{Determined from $L_*$ and \Teff.}
\tablenotetext{b}{Determined from $L_*$ and \Teff\ using Siess et al. (2000)
evolutionary tracks.}
\tablenotetext{c}{Upper limit on $P_{rot}$ determined from stellar radius and
measured \vsini.}
\tablenotetext{d}{Using Br-$\gamma$ line luminosity with Muzerolle
et al. (1998) relationship to determine $L_{acc}$ which is used with equation
(8) of Gullbring et al. (1998) assuming the disk truncation radius is equal
to $5 R_*$ to find $\dot M$.}
\tablenotetext{e}{Equatorial field strength for an assumed dipole using
magnetospheric accretion model of K\"onigl (1991).}
\tablenotetext{f}{Equatorial field strength for an assumed dipole using
magnetospheric accretion model of Cameron \& Campbell (1993).}
\tablenotetext{g}{Equatorial field strength for an assumed dipole using
magnetospheric accretion model of Shu et al. (1994).}
\tablenotetext{h}{Predicted X-ray luminosity based on the Pevtsov et al. 
(2003) relationship between $L_X$ and magnetic flux.}
\tablenotetext{i}{Computed with the NextGen model with \logg$=3.5$ which is
the closest in the grid to the predicted gravity.}
\tablenotetext{j}{Computed with the NextGen model with \logg$=4.0$ which is
the closest in the grid to the predicted gravity.}
\tablenotetext{k}{Same entries as line 1 except for a higher mass accretion
rate more typical of Class I sources.}
\end{deluxetable}

\end{document}